\documentclass[%
 reprint,
 amsmath,amssymb,
 aps,
 prc,
]{revtex4-2}

\usepackage{graphicx}
\usepackage{dcolumn}
\usepackage{bm}
\usepackage{braket}
\usepackage{xcolor}
\usepackage{amsmath}
\usepackage{ulem}
\usepackage{cancel}
\usepackage{booktabs}
\usepackage{multirow}

\newcommand{\vect}{\boldsymbol}

\usepackage{hyperref}

\begin{document}

\preprint{APS/123-QED}

\title{Spin--orbital entanglement and spatial anisotropy of deuteron}

\author{Ryota Tateishi}
\email{tateishi.ryouta.015@s.kyushu-u.ac.jp}
\affiliation{%
 Department of Physics, Kyushu University, Fukuoka 819-0395, Japan
}%
\author{Tokuro Fukui}
\email{tokuro.fukui@artsci.kyushu-u.ac.jp}
\affiliation{
 Faculty of Arts and Science, Kyushu University, Fukuoka 819-0395, Japan
}%
\affiliation{
 RIKEN Nishina Center, Wako 351-0198, Japan
}%

\date{\today}

\begin{abstract}
 \edef\oldrightskip{\the\rightskip}
\begin{description}
 \rightskip\oldrightskip\relax
 \item[Background]
 The nuclear tensor force couples
 the spin and orbital degrees of freedom, thereby generating spin--orbital (SO) entanglement.
 This interaction also induces spatial anisotropy of density distributions. 
 While the tensor force underlies both phenomena, the direct connection between them remains unclear.

 \item[Purpose]
 We clarify the quantitative relationship between the SO entanglement and spatial anisotropy of the deuteron.
 This system provides a fully analytical description of both features.

 \item[Methods]
 We quantify the SO entanglement by the entanglement entropy and spatial anisotropy by the Kullback--Leibler (KL) divergence. Their evolution is traced by continuously varying the $S$--$D$ coupling in the deuteron.

 \item[Results] 
 We derive analytical expressions for the entanglement entropy and KL divergence, both of which depend on $M$, the projection of the deuteron total angular momentum $J=1$.
 The stronger the tensor-force-induced $S$--$D$ coupling is, 
 the larger both the entanglement entropy and KL divergence become.

 \item[Conclusions]
 The analytical results and the positive correlation between the two quantities clearly demonstrate that the two features stem from the same underlying coupling mechanism governed by the algebraic structure of the deuteron wave function.

\end{description}
\end{abstract}

\maketitle


\section{Introduction}
\label{sec:intro}
A characteristic feature of interacting quantum systems is the coupling between different degrees of freedom. Such couplings generate quantum correlations, including entanglement, which characterize the internal structure of quantum states. This correlated structure is reflected in a wide variety of physical properties and phenomena.

Turning to nuclear systems, this general picture is particularly evident in the distinct operator components of nuclear forces. Among them, the tensor force directly couples the spin and orbital degrees of freedom. The deuteron provides a particularly transparent system for examining the consequences of this coupling. In its ground state, the tensor force mixes the \(S\)- and \(D\)-wave components, leaving a direct imprint of the spin--orbital (SO) entanglement on the wave function.

Here, the SO entanglement is defined with respect to a bipartite Hilbert space spanned by the spin and orbital components, ${\mathcal{H}_{\mathrm{spin}}\otimes\mathcal{H}_{\mathrm{orbit}}}$. 
Entanglement between spin and orbital degrees of freedom, including the same bipartition considered here, has long been studied in condensed-matter systems~\cite{PhysRevLett.96.147205,PhysRevB.75.195113,CHEN20101393,PhysRevB.86.224422,You_2015,PhysRevResearch.2.013353}. These studies have shown that interactions can strongly entangle spin and orbital sectors and that such entanglement is closely related to quantum phases, their transitions, and low-energy excitations. 

In nuclear systems, recent studies have explored nuclear forces from the perspective of quantum entanglement. 
In particular, the entanglement suppression within spin states of few-hadron systems---particularly two-nucleon systems---has been investigated extensively~\cite{PhysRevLett.122.102001,PhysRevC.107.025204,Kirchner2024,PhysRevD.110.014001,x52w-4rbs,mlt1-z7t2}. 
Initially identified in the framework of low-energy $S$-wave contact interactions~\cite{PhysRevLett.122.102001}, this concept has recently been examined in the presence of noncentral forces, with particular attention to the tensor force~\cite{mlt1-z7t2}. 
Nevertheless, these studies have primarily concerned entanglement among particle spins, rather than entanglement between the spin and orbital sectors of a nuclear state.

In addition to generating the SO entanglement, the tensor force also affects the spatial structure of nuclei.
Forest \textit{et al.}~\cite{Forest1996} demonstrated that the tensor force induces pronounced spatial anisotropy in light nuclei including the deuteron, leading to dumbbell-shaped and toroidal density distributions depending on the projection \(M\) of the total angular momentum.
Indeed, a more recent study~\cite{kpl8-9nyx} extended the structural picture of the deuteron by reformulating it in terms of principal stresses, force distributions, and torsional spin reorientation, thereby deepening our understanding from its static shape to its internal mechanical response.
Both the SO entanglement and spatial anisotropy originate from the tensor force, suggesting that the two phenomena may be closely related.
However, their quantitative relationship remains unclear.


In this work, we elucidate the relation between the SO entanglement and the spatial anisotropy of the deuteron. 
The deuteron is chosen as an ideal baseline for this exploration since it offers the significant advantage of allowing both the entanglement entropy and the Kullback--Leibler (KL) divergence, which respectively quantify the entanglement and anisotropy, to be calculated analytically. 
Furthermore, for the deuteron, the fixed isospin state factorizes from the SO sector and therefore does not contribute to the SO entanglement.
By continuously varying the $S$--$D$ coupling in the deuteron, we reveal the detailed mechanism behind the evolution of the correlation between the entanglement entropy and KL divergence.

This paper is organized as follows.
In Sec.~\ref{sec:form}, we present the formalism and method used in this work, including the entanglement entropy, KL divergence, and how to control the \(S\)-\(D\) coupling in the deuteron. Section~\ref{sec:res} is devoted to show the results and discuss the correlation between the SO entanglement and the spatial anisotropy of the deuteron. Finally, in Sec.~\ref{sec:concl}, we summarize our conclusions and discuss future perspectives.

 
\section{Formalism and Methods}
\label{sec:form}
\subsection{Spin--orbital entanglement entropy}
\label{sec:form_SO}
The ground state of the deuteron is expressed by the superposition of the $S$- and $D$-wave components as
 \begin{gather}
    \Ket{\Phi_M} 
    = 
    \left[|C_0|\Ket{[\psi_0\otimes\xi_1]_{1M}}
    +|C_2|e^{i\delta}\Ket{[\psi_2\otimes\xi_1]_{1M}}\right]\Ket{\chi_{00}},
    \label{eq:deuteron_SD}\\
    \Ket{[\psi_L\otimes\xi_S]_{JM}}
    = \sum_{M_LM_S}(LM_LSM_S|JM)\Ket{\psi_{LM_L}\xi_{SM_S}},
    \label{eq:LS_coupling}
 \end{gather}
where $\Ket{\psi_{LM_L}}$ denotes the orbital part with the orbital angular momentum \(L\) and its third component $M_L$, while $\Ket{\xi_{SM_S}}$ represents the spin-triplet part with \(S=1\) and ${M_S = -1,0,1}$.
They form orthonormal sets, satisfying ${\Braket{\psi_{LM_L}|\psi_{L'M_L'}}=\delta_{LL'}\delta_{M_LM_L'}}$ and ${\Braket{\xi_{SM_S}|\xi_{S'M_S'}}=\delta_{SS'}\delta_{M_SM_S'}}$.
The Clebsh–Gordan coefficients are expressed by $(\cdot\cdot\cdot\cdot|\cdot\cdot)$.
As the deuteron has the total angular momentum $J=1$, its projection $M$ runs over $-1$, $0$, and $1$.
The isospin-singlet part $\Ket{\chi_{TM_T}}$ with $T=M_T=0$ is factored out.
The coefficients $|C_0|$ and $|C_2|$ determine the $S$- and $D$-state probabilities expressed by $P_S=|C_0|^2$ and $P_D=|C_2|^2$, respectively. 
Then, the normalization condition reads \(P_S+P_D=1\).
The relative phase $\delta$ between the $S$- and $D$-wave components does not contribute to the density matrices and entanglement entropy provided that an appropriate basis including $\delta$ is chosen (see Table~\ref{tab:coefficients}).

Equations~\eqref{eq:deuteron_SD} and~\eqref{eq:LS_coupling} express the deuteron state $\Ket{\Phi_M}$ in terms of the orthonormal product basis \(\Ket{\psi_{L M_L}}\Ket{\xi_{1M_S}}\), by which we define the bipartition between the orbital and spin degrees of freedom. 
The entanglement associated with this bipartition is referred to as the
SO entanglement.
The isospin part of the deuteron is fixed to the \(T=0\) configuration and
is not involved in this bipartition.
We therefore trace out the isospin degree of freedom [see Eq.~\eqref{eq:density_matrix}].

For each projection \(M\), the deuteron state is expressed in a
four-dimensional product basis.
We collect the corresponding amplitudes into the coefficient vector $\mathbf{c}_M$,
\begin{align}
    \mathbf{c}_M
    =
    \begin{pmatrix}
        \alpha_M \\
        \beta_M \\
        \gamma_M \\
        \delta_M
    \end{pmatrix}.
    \label{eq:coefficient_vector}
\end{align}
The explicit values of $\alpha_M$, $\beta_M$, $\gamma_M$, and $\delta_M$, the corresponding product basis also, for each projection \(M\) are listed in
Table~\ref{tab:coefficients}.

\begin{table}[!t]
\caption{Coefficient vector components and basis vectors for each projection $M$.}
\label{tab:coefficients}
\centering
\begin{tabular*}{\columnwidth}{c@{\extracolsep{\fill}}c@{\extracolsep{\fill}}c@{\extracolsep{\fill}}c@{\extracolsep{\fill}}c@{\extracolsep{\fill}}c}
\toprule
 $M$ & $\begin{pmatrix}\alpha_M \\ \beta_M \\ \gamma_M \\       \delta_M \end{pmatrix}$ & Product basis  \\[0.7em]
\midrule
 $1$ & $\begin{pmatrix}|C_0|\\[0.7em] \sqrt{\dfrac{3}{5}}|C_2|\\[0.7em] -\sqrt{\dfrac{3}{10}}|C_2|\\[0.7em] \dfrac{1}{\sqrt{10}}|C_2| \end{pmatrix}$ & $\begin{pmatrix}\Ket{\psi_{00}\xi_{11}}\\[0.7em] e^{i\delta}\Ket{\psi_{22}\xi_{1,-1}}\\[0.7em] e^{i\delta}\Ket{\psi_{21}\xi_{10}}\\[0.7em] e^{i\delta}\Ket{\psi_{20}\xi_{11}} \end{pmatrix}$\\[0.7em]
\midrule
 $0$ & $\begin{pmatrix} |C_0|\\[0.7em] \sqrt{\dfrac{3}{10}}|C_2|\\[0.7em] -\sqrt{\dfrac{2}{5}}|C_2|\\[0.7em] \sqrt{\dfrac{3}{10}}|C_2| \end{pmatrix}$ & $\begin{pmatrix}\Ket{\psi_{00}\xi_{10}}\\[0.7em] e^{i\delta}\Ket{\psi_{21}\xi_{1,-1}}\\[0.7em] e^{i\delta}\Ket{\psi_{20}\xi_{10}}\\[0.7em] e^{i\delta}\Ket{\psi_{2,-1}\xi_{11}} \end{pmatrix}$\\[0.7em]
\midrule
 $-1$ & $\begin{pmatrix}|C_0|\\[0.7em] \dfrac{1}{\sqrt{10}}|C_2|\\[0.7em] -\sqrt{\dfrac{3}{10}}|C_2|\\[0.7em] \sqrt{\dfrac{3}{5}}|C_2| \end{pmatrix}$ & $\begin{pmatrix}\Ket{\psi_{00}\xi_{1,-1}}\\[0.7em] e^{i\delta}\Ket{\psi_{20}\xi_{1,-1}}\\[0.7em] e^{i\delta}\Ket{\psi_{2,-1}\xi_{10}}\\[0.7em] e^{i\delta}\Ket{\psi_{2,-2}\xi_{11}} \end{pmatrix}$\\
 \bottomrule
\end{tabular*}
\end{table}

The corresponding density matrix is then written in terms of the coefficient
vector as
\begin{align}
    \hat{\rho}_M
    &=
    \operatorname{Tr}_{\mathrm{isospin}}
    \left[
    \Ket{\Phi_M}
    \Bra{\Phi_M}
    \right]
    \notag \\
    &=
    \mathbf{c}_M
    \left(\mathbf{c}_M\right)^\dagger.
    \label{eq:density_matrix}
\end{align}
To evaluate the SO entanglement, we use the
reduced density matrix obtained by tracing over the orbital degree of freedom,
\begin{align}
    \hat{\rho}_M^{(\mathrm{spin})}
    =
    \operatorname{Tr}_{\mathrm{orbital}} \hat{\rho}_M,
    \label{eq:reduced_density_spin}
\end{align}
which is a $3 \times 3$ matrix.
Since $\hat{\rho}_M$ is constructed from the pure state, Eq.~\eqref{eq:deuteron_SD}, the reduced density matrices obtained by tracing over either sector have the same nonzero eigenvalues. 
We denote these eigenvalues by $\lambda_i^{(M)}$.
Thus, we define the SO entanglement entropy, i.e., the von Neumann entropy for the pure state $\Ket{\Phi_M}$, as
\begin{align}
    S_M
    &=
    -\operatorname{Tr}
    \left[
    \hat{\rho}_M^{(\mathrm{spin})}
    \ln \hat{\rho}_M^{(\mathrm{spin})}
    \right]
    \notag \\
    &=
    -\sum_i \lambda_i^{(M)}\ln\lambda_i^{(M)}.
    \label{eq:spin_orbital_entropy}
\end{align}

Since the reduced density matrix $\hat{\rho}_M^{(\mathrm{spin})}$ and its eigenvalues are expressed in terms of the coefficient vector $\mathbf{c}_M$, $S_M$ can also be obtained analytically as
\begin{align}
    S_{M=\pm1}
    &= -\left(1 - \frac{9}{10} \left| C_2 \right|^2\right) \ln \left(1 - \frac{9}{10} \left| C_2 \right|^2 \right)\notag \\
    &\quad - \frac{3}{10} \left| C_2 \right|^2 \ln \left( \frac{3}{10} \left| C_2 \right|^2 \right)\notag \\
    &\quad - \frac{3}{5} \left| C_2 \right|^2 \ln \left( \frac{3}{5} \left| C_2 \right|^2 \right),
    \label{SM_pm1}\\
    S_{M=0}
    &= -\left(1 - \frac{3}{5} \left| C_2 \right|^2\right) \ln \left(1 - \frac{3}{5} \left| C_2 \right|^2\right)\notag \\
       &\quad - \frac{3}{5} \left| C_2 \right|^2 \ln \left( \frac{3}{10} \left| C_2 \right|^2 \right).
    \label{SM_0}
\end{align}
The entanglement entropy $S_M$ depends solely on the \(D\)-state probability \(P_D\).
Equations~\eqref{SM_pm1} and~\eqref{SM_0} can also be derived from the Schmidt decomposition, as shown in Appendix~\ref{sec:Schmidt}.

\begin{figure}[!t]
\begin{center}
\includegraphics[width=0.5\textwidth]{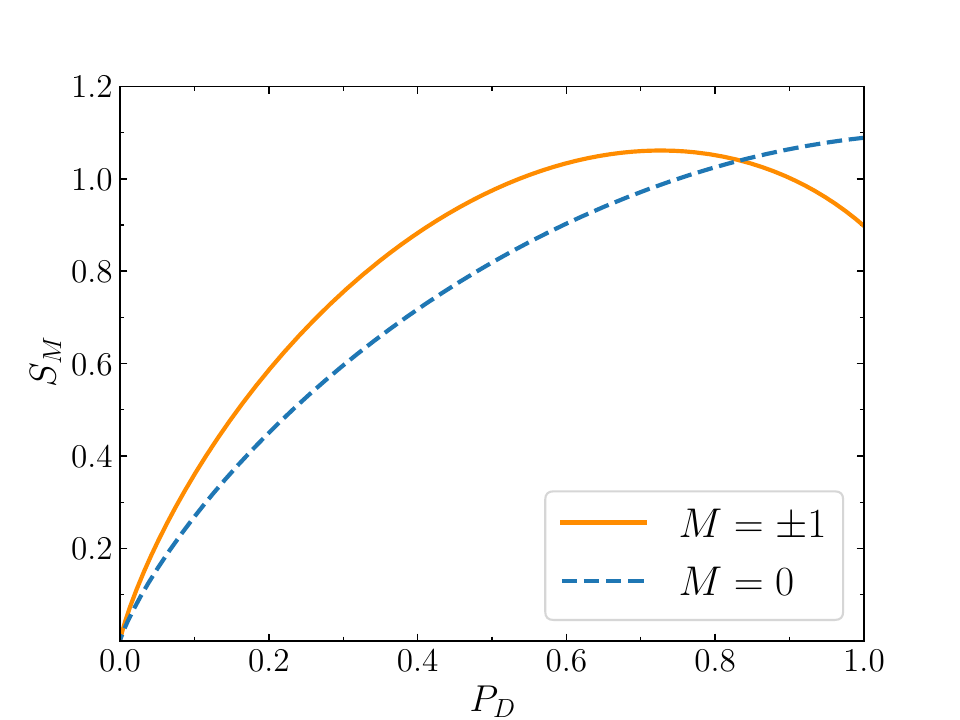}
\caption{SO entanglement entropy \(S_M\) as a function of the $D$-state probability for the spin projections $M=\pm1$ (solid line) and $M=0$ (dashed line) of the deuteron.
}
\label{fig:entropy}
\end{center}
\end{figure}

Figure~\ref{fig:entropy} shows $S_M$ as a function of $P_D$.
The entanglement entropy reaches its maximum value; ${S_{M=\pm1} \sim 1.061}$ at ${P_D \sim 0.727}$ and ${S_{M=0} \sim 1.089}$ at ${P_D=1}$.
Once $P_D$ is fixed, we can estimate the degree of the SO entanglement relative to the maximum value of $S_M$.

We add comments on a different bipartite Hilbert space 
spanned by the $S$- and $D$-wave components,
${\mathcal{H}_S\otimes\mathcal{H}_D}$, in the deuteron.
The corresponding entanglement entropy is the binary entropy defined by
\begin{align}
    H_{SD} = -|C_2|^2 \ln |C_2|^2
             -\left(1-|C_2|^2\right) \ln \left(1-|C_2|^2\right),
\end{align}
which quantifies the entanglement between the $S$- and $D$-state 
components.
Although $H_{SD}$ measures the entanglement originating from the tensor force, its behavior as a function of $P_D$ is rather trivial.
Moreover, because it is independent of $M$, this bipartition is unrelated to the spatial anisotropy considered in this paper.
Therefore, further investigation into this specific entanglement offers no additional physical insights within our current framework.


\subsection{Kullback--Leibler divergence}
\label{sec:form_KL}
We quantify the anisotropy of the deuteron density using the KL divergence, which measures the deviation of the density distributions from isotropy.
This allows us to characterize angular anisotropy that complements conventional deformation parameters~\cite{BohrDeformation,PhysRev.89.1102}.
To define the KL divergence, we first introduce the one-body density relative to the deuteron center of mass~\cite{Forest1996};
\begin{align}
    \tilde \rho_M(\vect r')
    =
    16 \Braket{2\vect r' \left| \operatorname{Tr}_{\mathrm{spin,\,isospin}}\hat\rho_M\right|2\vect r'}
\end{align}
where $\vect r'$ denotes the distance of a nucleon from the center of mass of the deuteron. Hence, the proton--neutron relative distance is defined by $\vect r = 2\vect r'$.
For each $M$, $\tilde\rho_M$ is given by
\begin{align}
 \tilde\rho_{M=\pm1}(\vect r') 
 &= 
 \frac{4}{\pi}[F_0(2r')+F_2(2r')P_2(\cos\theta))],
 \label{eq:density_M1}
 \\
 \tilde\rho_{M=0}(\vect r') 
 &= \frac{4}{\pi}[F_0(2r')-2F_2(2r')P_2(\cos\theta)],
 \label{eq:density_M0}
 \\
 F_0(r)
 &= \left[\phi_0(r)\right]^2+\left[\phi_2(r)\right]^2,
 \\
 F_2(r)
 &= \sqrt{2}\phi_0(r)\phi_2(r)-\frac{1}{2}\left[\phi_2(r)\right]^2.
\end{align}
The angular parts of $\vect r'$ are represented by ${\hat{\vect r} = \vect r/r = (\theta,\phi)}$ with the polar angle $\theta$ and azimuthal angle $\phi$. The density $\tilde\rho_M$ depends on the second-order Legendre polynomial $P_2$. We express the radial wave functions of the \(S\)- and \(D\)-wave components by $\phi_0$ and $\phi_2$, respectively. 
These wave functions are related to $\psi_{LM_L}$ involved in Eq.~\eqref{eq:deuteron_SD} through
\begin{align}
  |C_L|\psi_{LM_L}(\vect r) 
  &=
  \phi_L(r)Y_{LM_L}(\hat{\vect r}),
  \label{psi_phi}
\end{align}
where $Y_{LM_L}$ is the spherical Harmonics.
The normalization condition reads
\begin{align}
 \int_0^\infty dr r^2 \, \left[\phi_0(r)\right]^2 
 + \int_0^\infty dr r^2\, \left[\phi_2(r)\right]^2 \, =1.
 \label{PSPD1}
\end{align}
Note that, since the coupled radial Schr\"odinger equation for the deuteron is real, $\phi_0$ and $\phi_2$ can be chosen to be real. We adopt the phase convention in which the two components have the same sign and hence set $\delta=0$.


Next, we define the angular density distribution as
\begin{align}
 p_{M}(\hat{\vect r}) 
 &=
 \frac{1}{A}\int dr' r'^2 \tilde\rho_M(\vect r')
 \notag\\
 &= \frac{1}{4\pi}\left[1+B_{M}P_2(\cos\theta)\right],
 \label{pOmega}
\end{align}
where the mass number $A=2$ is introduced to ensure that
$p_M$ is normalized to be unity,
\begin{align}
  \int d\hat{\vect r} \, p_M(\hat{\vect r}) = 1.
  \label{norm_pOmega}
\end{align}
The coefficient $B_M$ is given by
\begin{align}
 B_{M=\pm1} &= -\frac{1}{2}P_D+\sqrt{2}I_{SD},
 \label{eq:B_M=pm1}\\
 B_{M=0} &= P_D-2\sqrt{2}I_{SD}.
  \label{eq:B_M=0}
\end{align}
The $D$-state probability $P_{D}$ and the $S$--$D$ interference integral $I_{SD}$ are calculated from
\begin{align}
 P_D &= \int_0^\infty dr \,r^2 \, \left[ \phi_2(r) \right]^2,
 \label{PD}\\
 I_{SD} &= \int_0^\infty dr \,r^2\, \phi_0(r)\phi_2(r).
\end{align}
Alternatively to Eq.~\eqref{pOmega}, $p_M$ can also be derived using the density matrix $\hat \rho_M$:
\begin{align}
    p_M(\hat{\vect r})
    =
    \Braket{\hat{\vect r}\left|\operatorname{Tr}_{r\sigma\tau} \hat \rho_M\right|\hat{\vect r}},
\end{align}
where the subscripts $r$, $\sigma$, and $\tau$ represent the radial, spin, and isospin components, respectively.

\begin{figure}[!t]
\begin{center}
\includegraphics[width=0.5\textwidth]{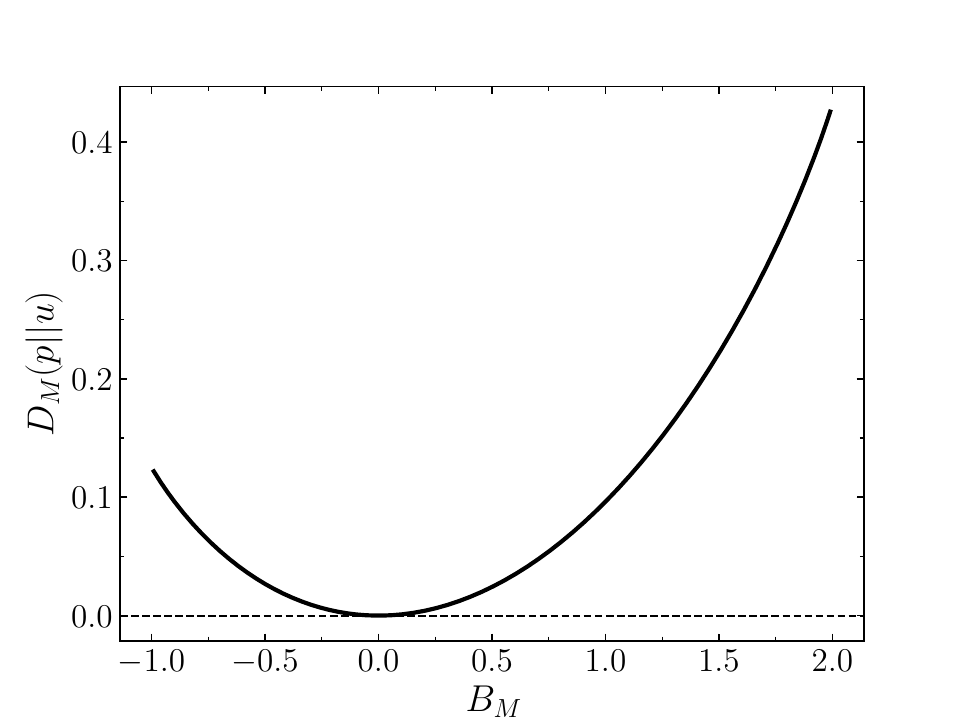}
\caption{The KL divergence $D_M$ as a function of \(B_M\).}
\label{fig:KLdivergence}
\end{center}
\end{figure}

Then, we can define the KL divergence as
\begin{align}
 D_{M}(p||u)
 &= \int d\hat{\vect r} \,   p_M(\hat{\vect r})\log{\frac{p_M(\hat{\vect r})}{u(\hat{\vect r})}}
 \notag\\
 &= \frac13\left[3\ln(B_M+1)+B_M-4\right]
 \notag\\
 &+\frac29\sqrt{\frac{3(2-B_M)^3}{B_M}}\arctan\left(\sqrt{\frac{3B_M}{2-B_M}}\right),
 \label{DKL}
\end{align}
where, as the reference distribution, we also introduce the isotropic angular density distribution, defined as
\begin{align}
 u(\hat{\vect r}) = \frac{1}{4\pi}.
 \label{uOmega}
\end{align}
The integration over $\hat{\vect r}$ in Eq.~\eqref{DKL} is evaluated using Mathematica~\cite{Mathematica}.

As expressed by Eq.~\eqref{DKL}, the KL divergence $D_M$ is a function of $B_M$. 
To avoid the singularities at ${B_M=-1}$ and $2$ in $D_M$, the range of $B_M$ is restricted to ${-1 < B_M < 2}$.
Furthermore, since $B_M$ depends on $P_D$ and $I_{SD}$, the domain of $D_M$ is implicitly constrained by the boundaries,
\begin{align}
    &0 \leq P_D \leq 1,
    \label{rangePD}\\
    &0 \leq I_{SD} \leq \sqrt{P_D(1-P_D)}.
    \label{rangeISD}  
\end{align}
In practice, we vary $P_D$ by controlling the $S$–$D$ coupling in solving the coupled-channel equations for the deuteron, as explained in Sec.~\ref{sec:form_SDscale}.
As a result, our calculations systematically yield ${I_{SD} > P_D}$ due to the $S$-wave dominance, which inherently restricts the coefficient to ${B_{M=\pm1}>0}$ and ${B_{M=0}<0}$.

Figure~\ref{fig:KLdivergence} shows the KL divergence $D_M$ as a function of \(B_M\). At \(B_M=0\), the angular density distribution $p_M$ does not depend on $\theta$, corresponding to the spherically symmetric density. Hence, ${D_M = 0}$ at ${B_M=0}$.
For \(B_M\neq0\), the Legendre polynomial $P_2$ in Eq.~\eqref{pOmega} induces an anisotropic density. 
This anisotropy is quantified by the deviation of $D_M$ from zero.

\subsection{Scaling $S$--$D$ coupling}
\label{sec:form_SDscale}
The deuteron wave function is obtained by solving the coupled-channel equations in momentum space~\cite{CDBonn2001}: 
\begin{alignat}{2}
  \tilde\phi_0(k)
  &= -\frac{\hat{M}}{\gamma^2+k^2}
     \int_0^\infty dk'\,k'^2 
  &&\bigl[V_{00}(k,k')\tilde\phi_0(k')
     \notag\\
  &&& +\lambda V_{02}(k,k')\tilde\phi_2(k')\bigr],
     \label{cceq0}\\
  \tilde\phi_2(k)
  &= -\frac{\hat{M}}{\gamma^2+k^2}
     \int_0^\infty dk'\,k'^2 
  &&\bigl[\lambda V_{20}(k,k')\tilde\phi_0(k')
     \notag\\
  &&& +V_{22}(k,k')\tilde\phi_2(k')\bigr].
     \label{cceq2}
\end{alignat}
Here, \(\tilde{\phi}_0\) and \(\tilde{\phi}_2\) are the
\(S\)- and \(D\)-wave radial functions in momentum space, respectively,
with $k$ the magnitude of the proton--neutron relative momentum.
The deuteron binding momentum $\gamma$ is associated with the $S$-matrix pole at $k = i\gamma$, and $\hat M$ is twice the proton--neutron reduced mass.
We employ the chiral nucleon--nucleon interaction at next-to-next-to-next-to-leading order (N$^3$LO)~\cite{PhysRevC.68.041001,MACHLEIDT20111} for the momentum-space potentials, $V_{00}$, $V_{02}$, $V_{20}$, and $V_{22}.$

To investigate how the channel coupling induced by the tensor force affects the SO entanglement and the spatial anisotropy of the deuteron, we scale the offdiagonal potentials $V_{02}$ and $V_{20}$ by introducing a scaling parameter $\lambda$.
Hence, calculations with $\lambda=1$ yield the realistic deuteron. While $\lambda > 1$ does not correspond to a realistic situation, varying this parameter allows us to elucidate the mechanism underlying the SO entanglement and deuteron anisotropy by revealing how far the realistic state is situated from ideal and/or extreme scenarios.

Note that calculations for $\lambda < 0$ are redundant since the physical quantities under consideration depend solely on $|\lambda|$.
More precisely, the sign inversion of the coupling potentials, $V_{02}$ and $V_{20}$, is equivalent to the unitary transformation,
\begin{align}
    \begin{pmatrix}
        V_{00} & -V_{02} \\
        -V_{20} & V_{22} \\
    \end{pmatrix}
    &=
    U
    \begin{pmatrix}
        V_{00} & V_{02} \\
        V_{20} & V_{22} \\
    \end{pmatrix}
    U^\dagger,
    \label{unitarytrans}
\end{align}
where the matrix $U$ is the third Pauli matrix,
\begin{align}
    U
    &=
    \begin{pmatrix}
        1 & 0 \\
        0 & -1 \\
    \end{pmatrix}.
    \label{unitarymat}
\end{align}
While applying $U$ to the deuteron state alters the relative phase between the $S$- and $D$-state components, a consistent unitary transformation of the relevant operators leaves the resulting physical quantities unchanged.
Consequently, $S_M$ and $D_M$ are independent on the sign of $\lambda$.

The coordinate-space \(S\)- and \(D\)-wave radial functions, which are defined by Eq.~\eqref{psi_phi}, are obtained via Fourier transform:
\begin{align}
\phi_L(r)
&=
\sqrt{\frac{2}{\pi}}
\int_{0}^{\infty} dk\, k^{2} j_L(kr)\,
\widetilde{\phi}_{L}(k),
\label{Fourier}
\end{align}
where \(j_{L}\) denotes the spherical Bessel function of order \(L\).

\section{Results and Discussion}
\label{sec:res}
\subsection{Spin--orbital entanglement in the deuteron}
\label{sec:res_entangle}
\begin{figure}[!t]
\begin{center}
\includegraphics[width=0.5\textwidth]{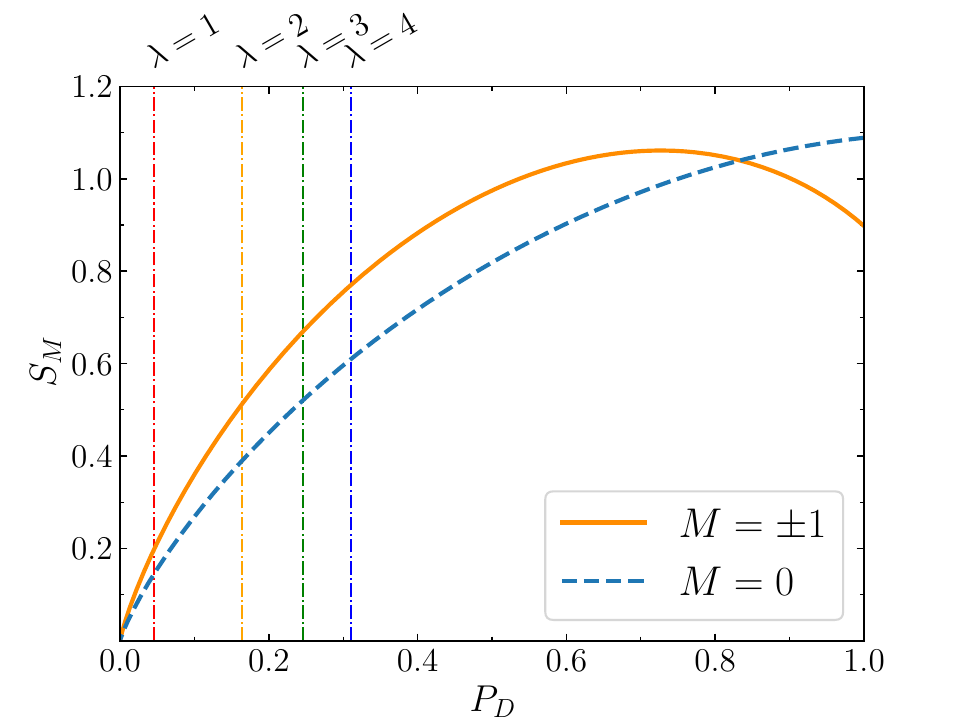}
\caption{Same as Fig.~\ref{fig:entropy}, but with vertical lines indicating \(P_D\) obtained with the scaling factor $\lambda$. 
}
\label{fig:entropy_lambda}
\end{center}
\end{figure}

Figure~\ref{fig:entropy_lambda} shows the SO entanglement entropy $S_M$ as a function of the \(D\)-state probability $P_D$ with the scaling factor \(\lambda\). 
The vertical lines indicate \(P_D\) when $\lambda$ takes 1, 2, 3 and 4. 

In the realistic case with $\lambda=1$, the chiral N$^3$LO interaction yields $P_D=0.0451$, resulting in ${S_{M=\pm1}=0.196}$ and ${S_{M=0}=0.143}$.
These values amount to $\sim 20$\% and $\sim 15$\% of their maximum values for $M=\pm1$ and $M=0$, respectively. Therefore, the SO entanglement of the realistic deuteron is rather limited.

With $\lambda\neq 1$, $P_D$ increases with $\lambda$ and eventually tends towards saturation. This saturation leads to the monotonic increase of \(S_M\).
Indeed, with $\lambda = 4$, $S_M$ reaches 71\% and 55\% of their maximum values for $M=\pm1$ and $M=0$, respectively.
These results indicate that a stronger \(S\)--\(D\) coupling, induced by the tensor force, generates greater quantum entanglement between the spin and orbital degrees of freedom in the deuteron.

The enhancement of the SO entanglement via the increased tensor force depends on $M$.
To further understand the physical implications of this \(M\)-dependence, we analyze the spatial anisotropy of the deuteron density in the next subsection.

\begin{figure*}[!t]
\begin{center}
\includegraphics[width=\textwidth]{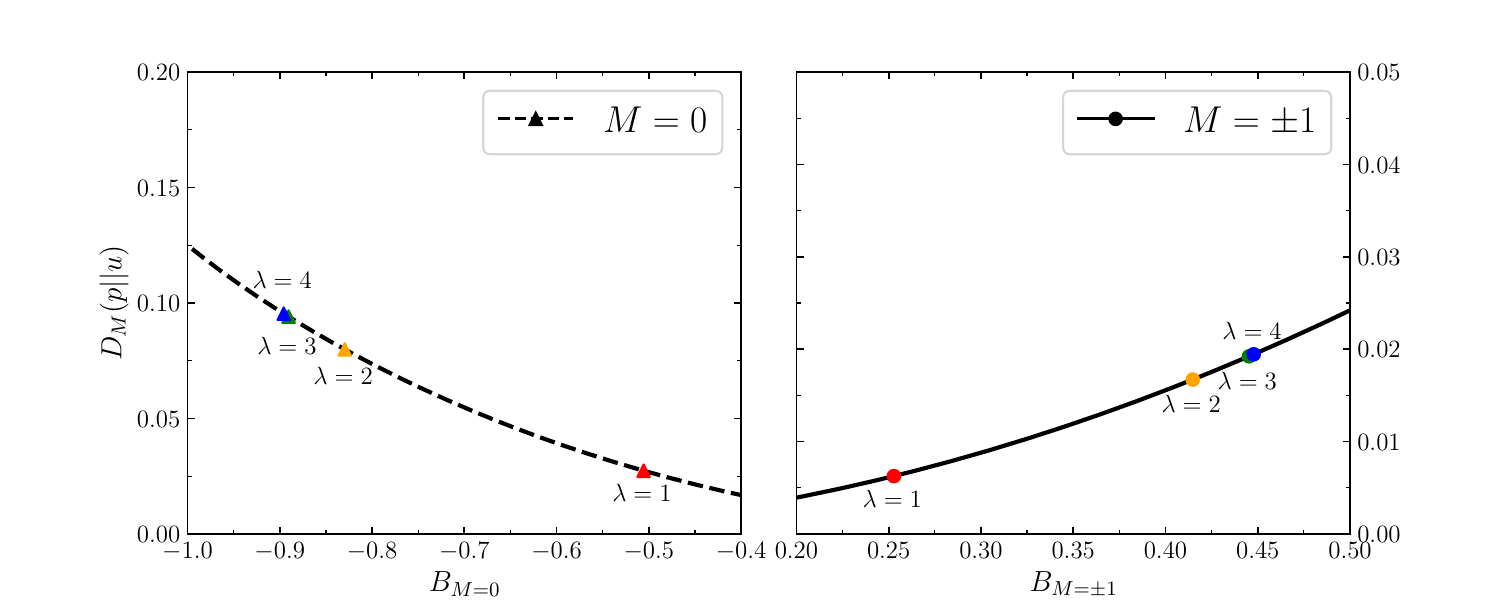}
\caption{Enlarged views of a part of Fig.~\ref{fig:KLdivergence}, with points plotted for different values of the scaling factor $\lambda$. Circles and triangles correspond to the spin projections $M=\pm1$ and $M=0$, respectively.}
\label{fig:KLdivergence_scale}
\end{center}
\end{figure*}

\subsection{Quantification of the deuteron anisotropy}
\label{sec:res_KL}
Figure~\ref{fig:KLdivergence_scale} shows the KL divergence $D_M$ with $\lambda=1$, 2, 3, and 4. 
The figure is divided into two panels according to the value of $M$.
In both cases, an increase in $\lambda$ drives the growth of $|B_M|$, which consequently enhances $D_M$.

Note that as $\lambda$ increases, the variation of $B_M$ becomes small in the range $1\leq\lambda\leq4$.
This behavior is consistent with the saturation of $P_D$, as discussed in Sec.~\ref{sec:res_entangle}.


The results in Fig.~\ref{fig:KLdivergence_scale} can be interpreted in terms of the tensor force.
Driven by the tensor force, the deuteron density forms dumbbell-shaped and toroidal distributions for $M=\pm1$ and $M=0$, respectively~\cite{Forest1996}.
Therefore, the stronger $S$--$D$ coupling further enhances the spatial anisotropy, leading to the increase in the deviation of $D_M$ from zero.
Accordingly, the dumbbell-shaped and toroidal density distributions develop as $\lambda$ increases. (see Fig.~\ref{fig:corr}).

The behavior of $D_M$ as a function $B_M$ is consistent with the description of nuclear deformation by the deformation parameters $\beta$ and $\gamma$~\cite{BohrDeformation,PhysRev.89.1102}.
For axially symmetric shapes, which are described by a signed deformation parameter $\beta$, with $\gamma$ fixed at zero, the prolate and oblate shapes correspond to $\beta>0$ and $\beta<0$, respectively.
Therefore, the coefficient $B_M$, which is the coefficient of the Legendre polynomial $P_2$ introduced in Eq.~\eqref{pOmega},
is qualitatively equivalent to $\beta$.

\subsection{Correlation between spin--orbital entanglement and spatial anisotropy}
\label{sec:res_corr}
\begin{figure*}[t]
\begin{center}
\includegraphics[width=\textwidth]{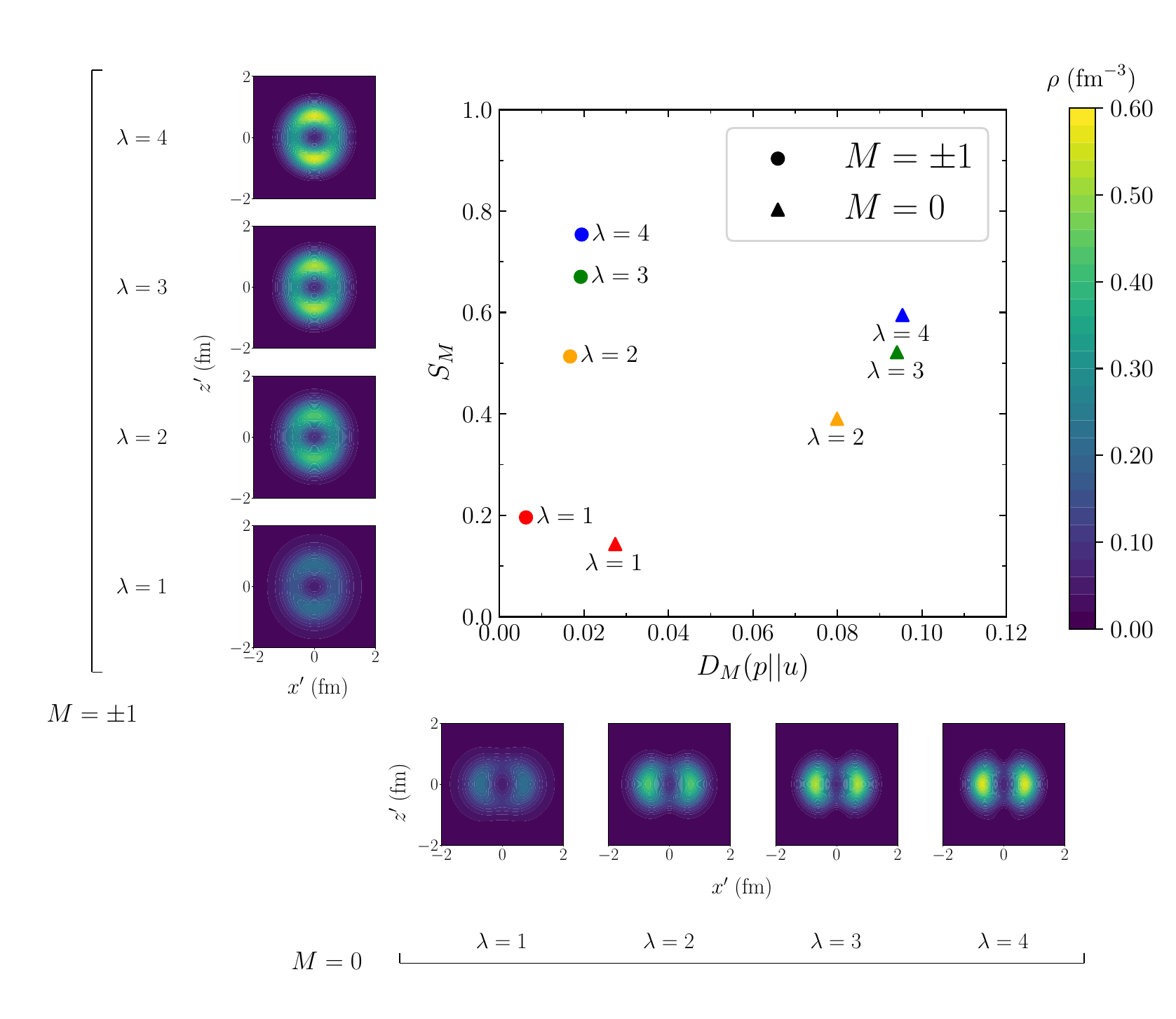}
\caption{Correlation between the SO entanglement entropy $S_M$ and the KL divergence $D_M$ for different values of the scaling factor $\lambda$. The circles and triangles correspond to the spin projections $M=\pm1$ and $M=0$, respectively.
The insets show the deuteron density associated with each symbol. The color scale represents the density.
Since the density distributions are independent of the azimuthal angle \(\phi\), they are represented in two dimensions, with the \(z\) axis taken as the quantization axis.
}
\label{fig:corr}
\end{center}
\end{figure*}

The growth of the SO entanglement entropy closely parallels the enhancement of the spatial anisotropy driven by the tensor force.
To clarify the physical mechanism behind this striking similarity, here we address the correlation between $S_M$ and $D_M$.

We show in Fig.~\ref{fig:corr} the correlation between $S_M$ and $D_M$ for $\lambda=1$, 2, 3, and 4.
As shown in Figs.~\ref{fig:entropy_lambda} and~\ref{fig:KLdivergence_scale}, both \(S_M\) and \(D_M\) increase with $\lambda$, leading to the positive correlation.

To understand these correlations intuitively, we also plot the density $\tilde\rho_M$ defined by Eqs.~\eqref{eq:density_M1} and~\eqref{eq:density_M0} as the insets associated with each symbol in Fig.~\ref{fig:corr}.
Since $\tilde\rho_M$ is symmetric with respect to the azimuthal angle $\phi$, we plot it in two dimensions with $\vect{r}'=(x',0,z')$.
The insets clearly demonstrate that $\tilde\rho_{M=\pm1}$ and $\tilde\rho_{M=0}$ respectively form the dumbbell-shaped and toroidal distributions.
Constrained by the normalization condition,
\begin{align}
    \int d\vect r' \tilde\rho_M(\vect r') = 2,
    \label{eq:density_norm}
\end{align}
the density exhibits a transition from a spatially broad distribution to a localized profile as $\lambda$ increases.

The geometric nature of the deuteron is inherently linked to the algebraic structure of its wave function.
Namely, the difference between the dumbbell-shaped and toroidal density distributions originates from the Clebsch--Gordan coefficients in Eq.~\eqref{eq:LS_coupling}, which also govern the characteristic behaviors of the SO entanglement entropy associated with $M$.
Consequently, both the SO entanglement and spatial anisotropy stem from the same underlying coupling mechanism.

\section{Conclusions and Perspectives}
\label{sec:concl}
We formulated the entanglement entropy $S_M$ to quantify the SO entanglement and proposed the KL divergence $D_M$ as a measure of the spatial anisotropy of the deuteron.
We then varied the strength of the $S$--$D$ coupling and examined the correlation between the SO entanglement entropy and KL divergence.

We found that the SO entanglement entropy grows as the $S$--$D$ coupling becomes stronger.
In addition, the KL divergence is enhanced with increasing the $S$--$D$ coupling, reflecting the development of the dumbbell-shaped and toroidal density distributions for $M=\pm1$ and $M=0$, respectively.
These results clearly demonstrate the positive correlation between \(S_M\) and \(D_M\).
The deuteron thus provides a particularly clean setting in which the tensor-induced coupling between spin and orbital degrees of freedom can be isolated without additional complexity from isospin, allowing its manifestations in the SO entanglement and spatial structure to be traced directly.

Our results reveal a connection between the quantum entanglement and geometric structure of the system, suggesting that the former may offer a useful perspective for understanding the nuclear force.
In the future, it would be interesting to examine whether this relation persists in heavier or more complex nuclei, where the isospin degrees of freedom need not to factorize from the SO sector.
In such systems, tracing out the isospin can yield a mixed SO state, for which mixed-state entanglement measures such as negativity 
could be useful for characterizing the SO entanglement.


\section*{Acknowledgments}
The authors thank R. Machleidt for supplying a numerical code for generating the chiral potentials in momentum space.
They also thank K. Ogata and F. Minato for valuable comments.
This work was supported by JSPS KAKENHI Grants No. 23KK0250 and NO. JP26K22335, RCNP COREnet Grant No. COREnet070, and JST ERATO Grant No. JPMJER2304.

\section*{data availability}
There are no publicly available research data or software supporting this manuscript. Requests for further information or data should be sent to the authors.

\appendix
\section{Schmidt decomposition}
\label{sec:Schmidt}
Since the deuteron ground state $\Ket{\Phi_M}$ given by Eq.~\eqref{eq:deuteron_SD} is a pure state, we can apply the Schmidt decomposition to it:
\begin{align}
    \Ket{\Phi_M} = \Ket{\chi_{00}} \sum_{i=1}^3 d_i^{(M)} \Ket{\tilde\psi_i^{(M)}} \otimes \Ket{\tilde\xi_i^{(M)}}.
    \label{PhiM_Schmidt}
\end{align}
Here, $\left\{\Ket{\tilde\psi_i^{(M)}}\right\}$ and $\left\{\Ket{\tilde\xi_i^{(M)}}\right\}$ are the orthonormal bases (ONBs) of the orbital and spin Hilbert spaces, respectively.
Since the Schmidt rank is three, both ONBs consist of three basis vectors. The ONBs and the Schmidt coefficient $d_i^{(M)} \geq 0$ can be explicitly written for each $M$, as summarized in Table~\ref{tab:Schmidt}.

The density matrix $\hat\rho_M$ defined by Eq.~\eqref{eq:density_matrix} is given by
\begin{align}
    \hat\rho_M
    &=
    \sum_{ij} d_i^{(M)} d_j^{(M)} 
    \Ket{\tilde\psi_i^{(M)}}\Bra{\tilde\psi_j^{(M)}}
    \otimes
    \Ket{\tilde\xi_i^{(M)}}\Bra{\tilde\xi_j^{(M)}}.
    \label{eq:density_matrix_Schmidt}
\end{align}
Then, the reduced density matrix $\hat\rho_M^{(\mathrm{spin})}$ is obtained from Eq.~\eqref{eq:reduced_density_spin} as
\begin{align}
    \hat\rho_M^{(\mathrm{spin})}
    &=
    \sum_{ij} d_i^{(M)} d_j^{(M)} 
    \notag\\
    &\times
    \operatorname{Tr}_{\mathrm{orbital}}\left[
    \Ket{\tilde\psi_i^{(M)}}\Bra{\tilde\psi_j^{(M)}}
    \right]
    \otimes
    \Ket{\tilde\xi_i^{(M)}}\Bra{\tilde\xi_j^{(M)}}
    \notag\\
    &=
    \sum_{i} \left(d_i^{(M)} \right)^2
    \Ket{\tilde\xi_i^{(M)}}\Bra{\tilde\xi_i^{(M)}}.
    \label{eq:reduced_density_spin_Schmidt}
\end{align}
The eigenvalues of Eq.~\eqref{eq:reduced_density_spin_Schmidt} coincide with the squared Schmidt coefficients, $\lambda_i^{(M)}=\left(d_i^{(M)} \right)^2$.
Applying Eq.~\eqref{eq:spin_orbital_entropy} with $d_i^{(M)}$ listed in Table~\ref{tab:Schmidt} yields the entanglement entropy $S_M$ identical to Eqs.~\eqref{SM_pm1} and~\eqref{SM_0}.

\begin{table*}[!t]
\caption{Schmidt coefficients $d_i^{(M)}$ and the associated ONBs: orbital state $\Ket{\tilde\psi_i^{(M)}}$ and spin $\Ket{\tilde\xi_i^{(M)}}$.
The $S$- and $D$-wave coefficients are $|C_0|$ and $|C_2|$, respectively.
The orbital states $\Ket{\psi_{LM_L}}$ and spin states $\Ket{\xi_{SM_S}}$ are introduced by Eqs.~\eqref{eq:deuteron_SD} and~\eqref{eq:LS_coupling}.
The relative phase $\delta$ originates from Eq.~\eqref{eq:deuteron_SD}.}
\label{tab:Schmidt}
\centering
\begin{tabular*}{\textwidth}{c@{\extracolsep{\fill}}c@{\extracolsep{\fill}}c@{\extracolsep{\fill}}c@{\extracolsep{\fill}}c}
\toprule
 $M$ & $i$ & $d_i^{(M)}$ & $\Ket{\tilde\psi_i^{(M)}}$ & $\Ket{\tilde\xi_i^{(M)}}$ \\
\midrule
 \multirow{3}{*}{$1$} & 1 & $\sqrt{1-\dfrac{9}{10}|C_2|^2}$ & $\left(1-\dfrac{9}{10}|C_2|^2\right)^{-1/2}\left(|C_0|\Ket{\psi_{00}}+\dfrac{1}{\sqrt{10}}|C_2|e^{i\delta}\Ket{\psi_{20}}\right)$ & $\Ket{\xi_{11}}$ \\[0.7em]
  & 2 & $\sqrt{\dfrac{3}{10}}|C_2|$ & $e^{i\left(\delta+\pi\right)}\Ket{\psi_{21}}$ & $\Ket{\xi_{10}}$ \\[0.7em]
  & 3 & $\sqrt{\dfrac{3}{5}}|C_2|$ & $e^{i\delta}\Ket{\psi_{22}}$ & $\Ket{\xi_{1,-1}}$ \\[0.7em]
\midrule
 \multirow{3}{*}{$0$} & 1 & $\sqrt{\dfrac{3}{10}}|C_2|$ & $e^{i\delta}\Ket{\psi_{2,-1}}$ & $\Ket{\xi_{11}}$ \\[0.7em]
  & 2 & $\sqrt{1-\dfrac{3}{5}|C_2|^2}$ & $\left(1-\dfrac{3}{5}|C_2|^2\right)^{-1/2}\left(|C_0|\Ket{\psi_{00}}-\sqrt{\dfrac25}|C_2|e^{i\delta}\Ket{\psi_{20}}\right)$ & $\Ket{\xi_{10}}$ \\[0.7em]
  & 3 & $\sqrt{\dfrac{3}{10}}|C_2|$ & $e^{i\delta}\Ket{\psi_{21}}$ & $\Ket{\xi_{1,-1}}$ \\[0.7em]
\midrule
 \multirow{3}{*}{$-1$} & 1 & $\sqrt{\dfrac{3}{5}}|C_2|$ & $e^{i\delta}\Ket{\psi_{2,-2}}$ & $\Ket{\xi_{11}}$ \\[0.7em]
  & 2 & $\sqrt{\dfrac{3}{10}}|C_2|$ & $e^{i\left(\delta+\pi\right)}\Ket{\psi_{2,-1}}$ & $\Ket{\xi_{10}}$ \\
  & 3 & $\sqrt{1-\dfrac{9}{10}|C_2|^2}$ & $\left(1-\dfrac{9}{10}|C_2|^2\right)^{-1/2}\left(|C_0|\Ket{\psi_{00}}+\dfrac{1}{\sqrt{10}}|C_2|e^{i\delta}\Ket{\psi_{20}}\right)$ & $\Ket{\xi_{1,-1}}$ \\
 \bottomrule
\end{tabular*}
\end{table*}

\bibliography{apssamp}

\end{document}